\documentclass[aps,prd,preprint,superscriptaddress,tightenlines,nofootinbib]{revtex4}

\usepackage{epsfig}
\usepackage{graphicx}
\usepackage{dcolumn}
\usepackage{bm}

\newcommand{\etal}{{\it et al.}}

\begin{document}

\preprint{\tighten\vbox{\hbox{\hfil CLEO CONF 06-17}
}}

\title{Measurement of  $D_s^+\to\mu^+\nu$ and the Decay Constant
$f_{D_s}$\hspace*{2mm}}
\thanks{Submitted to the 33$^{\rm rd}$ International Conference on High Energy
Physics, July 26 - August 2, 2006, Moscow}

\author{M.~Artuso}
\author{S.~Blusk}
\author{J.~Butt}
\author{J.~Li}
\author{N.~Menaa}
\author{R.~Mountain}
\author{S.~Nisar}
\author{K.~Randrianarivony}
\author{R.~Redjimi}
\author{R.~Sia}
\author{T.~Skwarnicki}
\author{S.~Stone}
\author{J.~C.~Wang}
\author{K.~Zhang}
\affiliation{Syracuse University, Syracuse, New York 13244}
% ...
\collaboration{CLEO Collaboration} 
\noaffiliation
\date{July 25, 2006}

\begin{abstract}
We examine $e^+e^-\to D_s^-D_s^{*+}$ or $D_s^{-*}D_s^{+}$
collisions at 4170 MeV using the CLEO-c detector in order to
measure the decay constant $f_{D_s^+}$. We use the $D_s^+\to
\ell^+\nu$ channel, where the $\ell^+$ designates either a $\mu^+$
or a $\tau^+$. Analyzing both modes simultaneously, we determine
${\cal{B}}(D_s^+\to \mu^+\nu)= (0.657\pm 0.090\pm0.028)$\%,
${\cal{B}}(D_s^+\to \tau^+\nu)= (7.1\pm 1.4\pm0.3)$\%, and extract
$f_{D_s^+}=282\pm 16 \pm 7 {~\rm MeV}$. Combining with our
previous determination of ${\cal{B}}(D^+\to \mu^+\nu)$, we find
that the ratio $f_{D_s^+}/f_{D^+}=1.27\pm 0.12\pm 0.03$. (All new
results here are preliminary.) We compare with current theoretical
estimates.
\end{abstract}


\maketitle \tighten

\section{Introduction}

To extract precise information on the size of
Cabibbo-Kobayashi-Maskawa matrix elements from $B-\overline{B}$
mixing measurements the ratio of ``leptonic decay constants,"
$f_i$ for $B_d$ and $B_s$ mesons must be well known
\cite{formula-mix}. Indeed, the recent measurement of
$B_s^0-\overline{B}_s^0$ mixing by CDF \cite{CDF} that can now be
compared to the very well measured $B^0$ mixing \cite{PDG}, has
pointed out the urgent need for precise numbers
\cite{Belle-taunu}. The $f_i$ have been calculated theoretically.
The most promising of these calculations are based on
lattice-gauge theory that include the light quark loops
\cite{Davies}. In order to ensure that these theories can
adequately predict $f_{B_s}/f_{B_d}$ it is useful to check the
analogous ratio from charm decays $f_{D^+_s}/f_{D^+}$. We have
previously measured $f_{D^+}$ \cite{our-fDp,DptomunPRD}. Here we
present the most precise measurement to date of $f_{D_s^+}$ and
the ratio $f_{D_s^+}/f_{D^+}$.

\begin{figure}[htbp]
 \vskip 0.00cm
 \centerline{ \epsfxsize=3.0in \epsffile{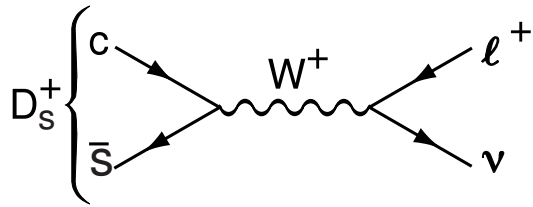} }
 \caption{The decay diagram for $D_s^+\to \ell^+\nu$.} \label{Dstomunu}
 \end{figure}

The only way in the Standard Model (SM) for a $D_s$ meson to decay
purely leptonically, via annihilation through a virtual $W^+$, is
shown in Fig.~\ref{Dstomunu}. The decay rate is given by
\cite{Formula1}
\begin{equation}
\Gamma(D_s^+\to \ell^+\nu) = {{G_F^2}\over
8\pi}f_{D_s^+}^2m_{\ell}^2M_{D_s^+} \left(1-{m_{\ell}^2\over
M_{D_s^+}^2}\right)^2 \left|V_{cs}\right|^2~~~, \label{eq:equ_rate}
\end{equation}
where $M_{D_s^+}$ is the $D_s^+$ mass, $m_{\ell}$ is the mass of the
charged final state lepton, $V_{cs}$ is a Cabibbo-Kobayashi-Maskawa
matrix element with a value we take equal to 0.9737 \cite{PDG}, and
$G_F$ is the Fermi coupling constant.

Previous measurements of $f_{D_s^+}$ have been hampered by a lack
of statistical precision, and relatively large systematic errors
\cite{PDG}. One large systematic error source has been the lack of
knowledge of the absolute branching ratio for $D_s^+\to\phi\pi^+$,
the mode which most measurements have used for normalization
\cite{stone-fpcp}. The results we report here will not have this
problem.

In this paper we analyze both $D_s\to\mu^+\nu$ and
$D_s\to\tau^+\nu$, $\tau^+\to \pi^+\overline{\nu}$. Both $D_s$
decays are helicity suppressed because the $D_s$ is a spin-0
particle, and the final state consists of a naturally left-handed
spin-1/2 neutrino and a naturally right-handed spin-1/2 anti-lepton
that have equal energies and opposite momenta. The ratio of decay
rates for any two different leptons is then fixed by well-known
masses. For example, for $\tau^+\nu$ to $\mu^+\nu$, the expected
ratio is

\begin{equation}
R\equiv \frac{\Gamma(D_s^+\to \tau^+\nu)}{\Gamma(D_s^+\to
\mu^+\nu)}= {{m_{\tau^+}^2 \left(1-{m_{\tau^+}^2\over
M_{D_s^+}^2}\right)^2}\over{m_{\mu^+}^2 \left(1-{m_{\mu^+}^2\over
M_{D_s^+}^2}\right)^2}}~~. \label{eq:tntomu}
\end{equation}
Using measured masses \cite{PDG}, this expression yields a value
of 9.72 with a negligibly small error. After multiplying by
${\cal{B}}(\tau^+\to\pi^+\overline{\nu}$ of 11.06\%, the ratio is
1.076 for $\tau^+\nu$ with respect to $\mu^+\nu$, when
$\tau^+\to\pi^+\overline{\nu}$.

\section{Experimental Method}
\subsection{Selection of $D_s$ Candidates}
The CLEO-c detector is equipped to measure the momenta and
directions of charged particles, identify them using specific
ionization (dE/dx) and Cherenkov light (RICH), detect photons and
determine their directions and energies \cite{CLEODR}.

In this study we use 200 pb$^{-1}$ of data produced in $e^+e^-$
collisions using the Cornell Electron Storage Ring (CESR) and
recorded near 4.170 GeV. Here the cross-section for
$D_s^{*+}D_s^-$+$D_s^{+}D_s^{*-}$ is $\sim$1 nb, with $D_s^+D_s^-$
production being only $\sim$5\% of this rate \cite{DsDsCBX}. $D$
mesons are also produced mostly as $D^{*}\overline{D^*}$, with a
cross-section of $\sim$5 nb, and in
$D^*\overline{D}+D\overline{D^*}$ final states with a cross-section
of $\sim$2 nb. The $D\overline{D}$ is a relatively small $\sim$0.2
nb \cite{poling}. There also appears to be $D\overline{D}$
production with extra pions. The underlying light quark ``continuum"
background is about 12 nb. The relatively large cross-sections,
relatively large branching ratios and sufficient luminosities, allow
us to fully reconstruct one $D_s$ as a ``tag," and examine the
properties of the other. In this paper we designate the tag as a
$D_s^-$ and examine the leptonic decays of the $D_s^+$, though in
reality we use both charges. Track selection, particle
identification, $\gamma$, $\pi^0$, $K_S$ and muon selection cuts are
identical to those described in Ref. \cite{our-fDp}.

\begin{figure}[htb]
\includegraphics[width=74mm]{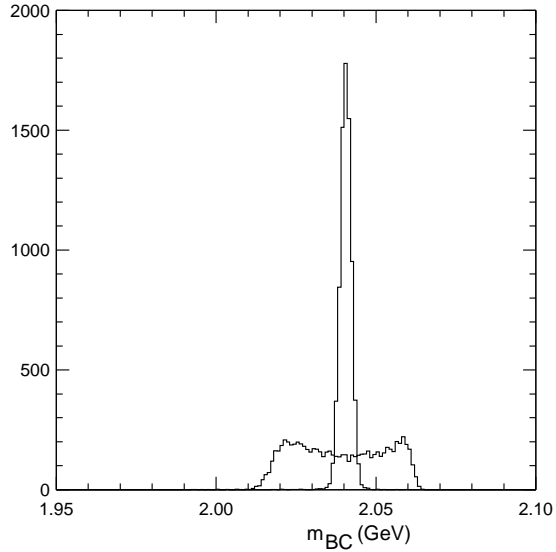}
\vspace{0.44mm}\caption{The beam constrained mass $m_{BC}$ from
Monte Carlo simulation of $e^+e^-\to D_s^+D_s^{*-}$,
$D_s^{\pm}\to\phi\pi^{\pm}$ at 4170 MeV. The narrow peak is from the
$D_s^+$ and the wider one from the $D_s^-$. (The distributions are
not centered at the $D_s^+$ or $D_s^{*+}$ masses, because the
reconstructed particles are assumed to have the energy of the
beam.)} \label{mbc}
\end{figure}

The events we use here occur when $e^+e^-\to D_s^{*+}D_s^-$ or
$D_s^{+}D_s^{*-}$. We will reconstruct tags from both final
states. The beam constrained mass, $m_{BC}$, is formed by using
the beam energy to construct the mass \cite{our-fDp}. If we do not
detect the photon and reconstruct the $m_{BC}$ distribution, we
obtain the distribution from Monte Carlo shown in Fig.~\ref{mbc}.
The narrow peak occurs when the reconstructed $D_s$ does not come
from the $D_s^*$ decay.

\begin{figure}[hbtp]
\centering
\includegraphics[width=5in]{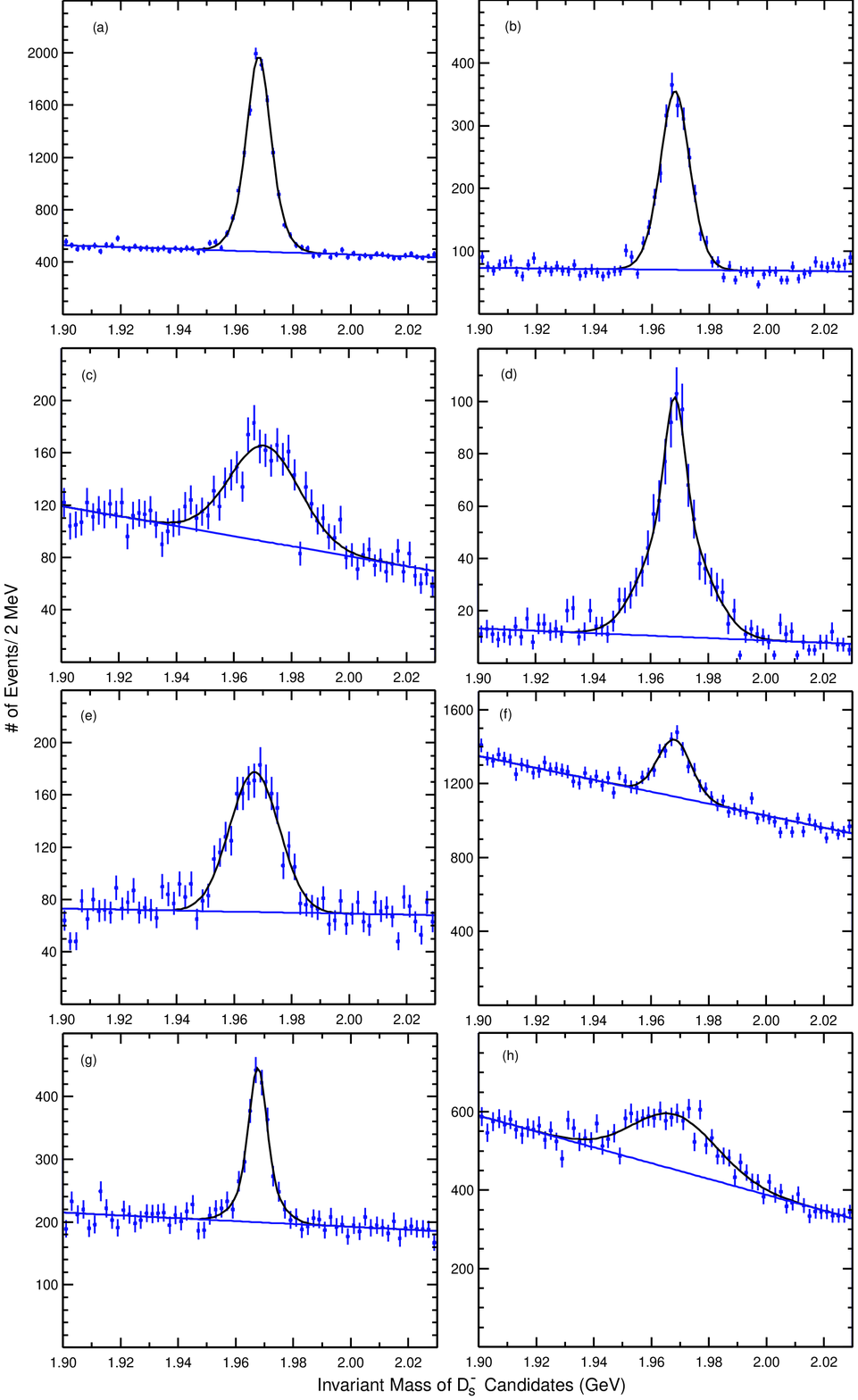}
\caption{Invariant mass of $D_s^-$ candidates (a) $K^+K^-\pi^+$,
(b) $K_SK^+$, (c) $\eta\pi^+$, (d) $\eta'\pi^+$, (e) $\phi\rho^+$,
(f) $\pi^+\pi^-\pi^+$, (g) $K^{*+}K^{*0}$,  (h) $\eta\rho^+$,
after requiring the total energy to be consistent with the beam
energy. The curves are fits to two-Gaussian signal functions plus
a linear background.
 } \label{Inv-mass}
\end{figure}

Rather than selecting events based on $m_{BC}$, we first select an
interval that accepts most of the events, $2.015 >m_{BC}>2.067$
GeV, and examine the invariant mass.  Distributions from data for
the 8 modes we use in this analysis are shown in
Fig.~\ref{Inv-mass}. Note that the resolution in invariant mass is
excellent, and the backgrounds not very large, at least in these
modes. To determine the number of tags we fit the invariant mass
distributions to the sum of two Gaussians centered at the $D_s^-$
mass. The r.m.s. resolution ($\sigma$) is defined as
\begin{equation}
\sigma = f_1\sigma_1+(1-f_1)\sigma_2, \label{eq:twoGauss}
\end{equation}
where $\sigma_1$ and $\sigma_2$ are the individual widths of the
two Gaussians and $f_1$ is the fractional area of the first
Gaussian. The number of tags in each mode is listed in
Table~\ref{tab:Ntags}.

\begin{table}[htb]
\begin{center}

\caption{Tagging modes and numbers of signal and background events,
within $\pm 2.5\sigma$ for all modes, except $\eta\rho^+$ ($\pm
2\sigma$), from two-Gaussian fits to the invariant mass plots.
\label{tab:Ntags}}
\begin{tabular}{lcc}
 \hline\hline
    Mode  &  \#            &  Background \\ \hline
$K^+K^-\pi^+ $ & $8446\pm160$   & 6792\\
$K_S K^+$ & 1852$\pm$62 & 1021\\
$\eta\pi^+$; $\eta\to\gamma\gamma$ & $1101\pm80$  & 2803\\
$\eta'\pi^+$; $\eta'\to\pi^+\pi^-\eta$, $\eta\to\gamma\gamma$ & 786$ \pm $37  &242 \\
$\phi\rho^+$; $\phi\to K^+K^-$, $\rho^+\to \pi^+\pi^0$ & 1140$ \pm $59  &1515 \\
$\pi^+\pi^-\pi^+$ & 2214$ \pm $156  & 15668\\
$K^{*+}K^{*0}$; $K^{*+}\to K_S\pi^+$, $K^{*0}\to K^-\pi^+$ & 1197$
\pm$81& 2955\\
$\eta\rho^+$; $\eta\to\gamma\gamma$, $\rho^+\to \pi^+\pi^0$ & 2449$ \pm $185  &13043 \\
\hline
Sum &  $19185\pm325 $ &44039 \\
\hline\hline
\end{tabular}
\end{center}
\end{table}

\subsection{Procedure for Finding Leptonic Decays}
In this analysis we will be selecting events from two processes
one where $D_s^+\to \mu^+\nu$ and the other when $D_s^+\to
\tau^+\nu$, $\tau^+\to\pi^+\overline{\nu}$.\footnote{In this paper
we use the charge conjugate mode in addition to the specified
charge mode.} We first have to select a sample of tag events. We
require the invariant masses, shown in Fig.~\ref{Inv-mass} to be
within $\pm ~2.5\sigma$ of the known $D_s^-$ mass (here $\sigma$
is the r.m.s. width). Then we look for an additional photon
candidate in the event that satisfies our shower shape
requirement. Regardless of whether or not the photon forms a
$D_s^*$ with the tag, for real $D_s^*D_s$ events, the missing mass
squared, MM$^{*2}$ recoiling against the photon and the $D_s^-$
tag should peak at the $D_s^+$ mass. We calculate
\begin{equation}
{\rm MM}^{*2}=\left(E_{\rm CM}-E_D-E_{\gamma}\right)^2-
\left(-\overrightarrow{p_D}-\overrightarrow{p_{\gamma}}\right)^2,
\end{equation}
where $E_{\rm CM}$ is the center of mass energy, $E_{D}$
($\overrightarrow{p_D}$) is the energy  (momentum) of the fully
reconstructed $D_s^-$ tag, $E_{\gamma}$
($\overrightarrow{p_{\gamma}}$) is the energy (momentum) of the
additional photon. In performing this calculation we use a kinematic
fit that constrains the decay products of the $D_s^-$  to the known
$D_s$ mass and conserves overall momentum and energy. All photons in
the event are used, except for those that are decay products of the
$D_s^-$ candidate.

\begin{figure}[hbt]
\centering
\includegraphics[width=4in]{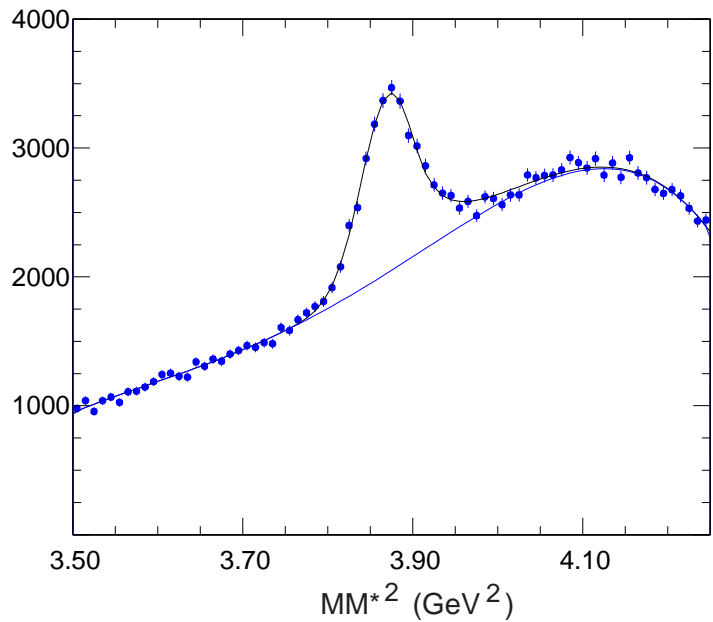}
\caption{The MM*$^2$ distribution from events with a photon in
addition to the $D_s^-$ tag. The curve is a fit to the Crystal Ball
function and a 5th order Chebychev background function.}
\label{MMstar2}
\end{figure}

\begin{figure}[hbt]
\centering
\includegraphics[width=4in]{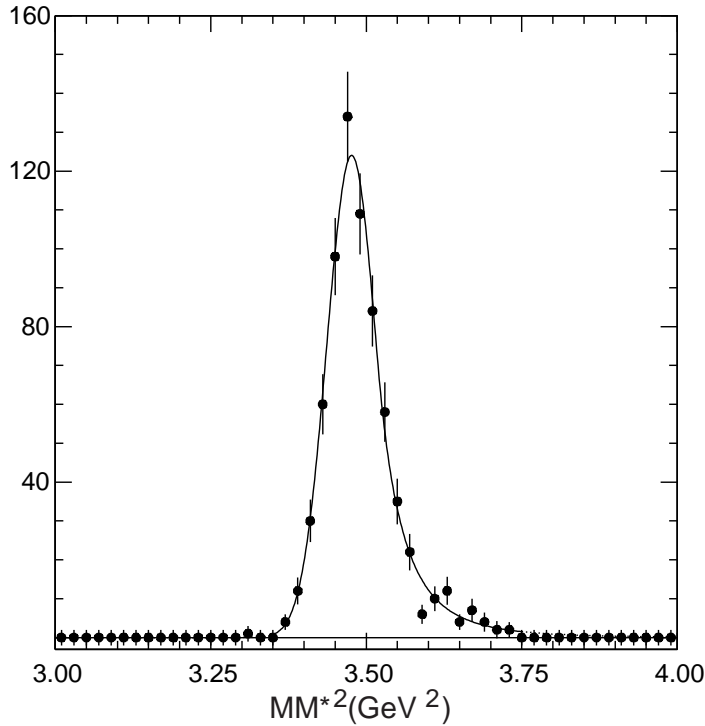}
\caption{The MM*$^2$ distribution from a sample of fully
reconstructed $D^{*0}D^0$ events where one $D^0$ is ignored. The
curve is a fit to the Crystal Ball function.}
\label{data-D0-MM2-DoubleTags}
\end{figure}

The MM$^{*2}$ from the $D_s^-$ tag sample data is shown in
Fig.~\ref{MMstar2}. We fit this distribution to determine the number
of tag events. This procedure is enhanced by having information on
the shape of the signal function. One possibility is to use the
Monte Carlo simulation for this purpose. Our relatively large sample
of fully reconstructed $D^{*0}D^0$ events allows us to examine the
signal shape in data when one neutral $D$ is ignored. This sample is
shown in Fig.~\ref{data-D0-MM2-DoubleTags}. The signal is fit to a
Crystal Ball function \cite{CBL,taunu}. The $\sigma$ parameter, that
represents the width of the distribution, is found to be
0.039$\pm$0.02 GeV$^2$ compared with the Monte Carlo estimate of
0.310$\pm$0.003 GeV$^2$. The Monte Carlo does not reproduce well the
width of the distribution, so we do not use it here. The energy of
photons from the $D^0$ and $D_s$ events are somewhat different due
to the different masses of the parent hadrons. Thus we cannot use
the $\sigma$ found here directly. We do, however, get an estimate of
the parameters of the tail of the distribution.

Using the fixed tail parameters of the Crystal Ball function, and
a 5th order Chebyshev polynomial background, we find 12604$\pm$423
signal events. After selecting events within an interval
$3.978>$MM$^{*2}>3.776$ GeV$^2$, we are left with 11880$\pm$399
events. A systematic error of 3\% is assigned by seeing how the
event yields vary when the width of the distribution is changed by
$\pm 1\sigma$ and an additional systematic error of 3\% is added
in quadrature from changing the form of the background function
and the fit range. There is also a small enhancement of 4\% on our
ability to find tags in $\mu^+\nu$ (or $\tau^+\nu$,
$\tau^+\to\pi^+\overline{\nu}$) events as compared with generic
events to which we assign a 10\% systematic error or $\pm$0.4\%,
again added in quadrature.

We next describe the search for $D_s^+\to\mu^+\nu$. Candidate events
are searched for by selecting events with only a single extra track
with opposite sign of charge to the tag, and also require that there
not be an extra neutral energy cluster in excess of 300 MeV. Since
here we are searching for events where there is a single missing
neutrino, the missing mass squared, MM$^2$, evaluated by taking into
account the seen $\mu^+$, $D_s^-$, and the $\gamma$ should peak at
zero; the MM$^2$ is computed as

\begin{equation}
\label{eq:mm2} {\rm MM}^2=\left(E_{\rm
CM}-E_{D}-E_{\gamma}-E_{\mu}\right)^2
           -\left(-\overrightarrow{p_ D}-\overrightarrow{p_{\gamma}}
           -\overrightarrow{p_{\mu}}\right)^2,
\end{equation}
where $E_{\rm CM}$ is the center of mass energy, $E_{D}$
($\overrightarrow{p_D}$) is the energy  (momentum) of the fully
reconstructed $D_s^-$ tag, $E_{\gamma}$
($\overrightarrow{p_{\gamma}}$) is the energy (momentum) of the
additional photon, and $E_{\mu}$ ($\overrightarrow{p_{\mu}}$) is
the energy (momentum) of the candidate muon track.

We also make use of a set of kinematical constraints and fit each
event to two hypotheses one of which is that the $D_s^-$ tag is the
daughter of a $D_s^{*-}$ and the other that the $D_s^{*+}$ decays
into $\gamma D_s^+$, with the $D_s^+$ subsequently decaying into
$\mu^+\nu$. The hypothesis with the lowest $\chi^2$ is kept. If
there is more than one photon candidate in an event we choose only
the lowest $\chi^2$ choice among all the candidates and hypotheses.

The kinematical constraints are
\begin{eqnarray}
\label{eq:constr}
&&\overrightarrow{p}_{\!D_s}+\overrightarrow{p}_{\!D_s^*}=0
\\\nonumber &&E_{\rm CM}=E_{D_s}+E_{D_s^*}\\\nonumber
&&E_{D_s^*}=\frac{E_{\rm
CM}}{2}+\frac{M_{D_s^*}^2-M_{D_s}^2}{2E_{\rm CM}}{\rm~or~}
E_{D_s}=\frac{E_{\rm CM}}{2}-\frac{M_{D_s^*}^2-M_{D_s}^2}{2E_{\rm
CM}}\\\nonumber &&M_{D_s^*}-M_{D_s}=143.6 {\rm ~MeV}.
\end{eqnarray}
In addition, we constrain the invariant mass of the $D_s^-$ tag to
the known $D_s$ mass. This gives us a total of 7 constraints. The
missing neutrino four-vector needs to be determined, so we are left
with a three-constraint fit. We preform a standard iterative fit
minimizing $\chi^2$. As we do not want to be subject to systematic
uncertainties that depending on understanding the absolute scale of
the errors, we do not make a $\chi^2$ cut but simply choose the
photon and the decay sequence in each event with the minimum
$\chi^2$.

\section{Signal Reconstruction}

In this analysis, we consider three separate cases: (i) the track
deposits $<$~300 MeV in the calorimeter, characteristic of a
non-interacting pion or a muon; (ii) the track deposits $>$~300 MeV
in the calorimeter, characteristic of an interacting pion; (iii) the
track satisfies our electron selection criteria defined below. Then
we separately study the MM$^2$ distributions for these three cases.
The separation between muons and pions is not unique. Case (i)
contains 99\% of the muons but also 60\% of the pions, while case
(ii) includes 1\% of the muons and 40\% of the pions
\cite{DptomunPRD}. Case (iii) does not include any signal but is
used later for background estimation.

We exclude events with more than one additional, opposite-sign
charged track in addition to the tagged $D_s^-$, or with extra
neutral energy. Specifically, we veto events with extra charged
tracks arising from the event vertex or having a maximum neutral
energy cluster, consistent with being a photon, of more than 300
MeV. These cuts are highly effective in reducing backgrounds. The
photon energy cut is especially useful to reject $D_s^+\to
\pi^+\pi^0$ decays, should this mode be significant, and $\eta\pi^+$
decays.

The track candidates are required to be within the barrel region of
the detector $|\cos\theta|<0.81$, where $\theta$ is the angle with
respect to the beam. For cases (i) and (ii) we insist that the track
not be identified as a kaon. For electron identification we require
a match between the momentum measurement in the tracking system and
the energy deposited in the CsI calorimeter and the shape of the
energy distribution among the crystals is consistent with that
expected for an electromagnetic shower.

\subsection{The Expected MM$^2$ Spectrum}
\begin{figure}[hbt]
\centering
\includegraphics[width=5in]{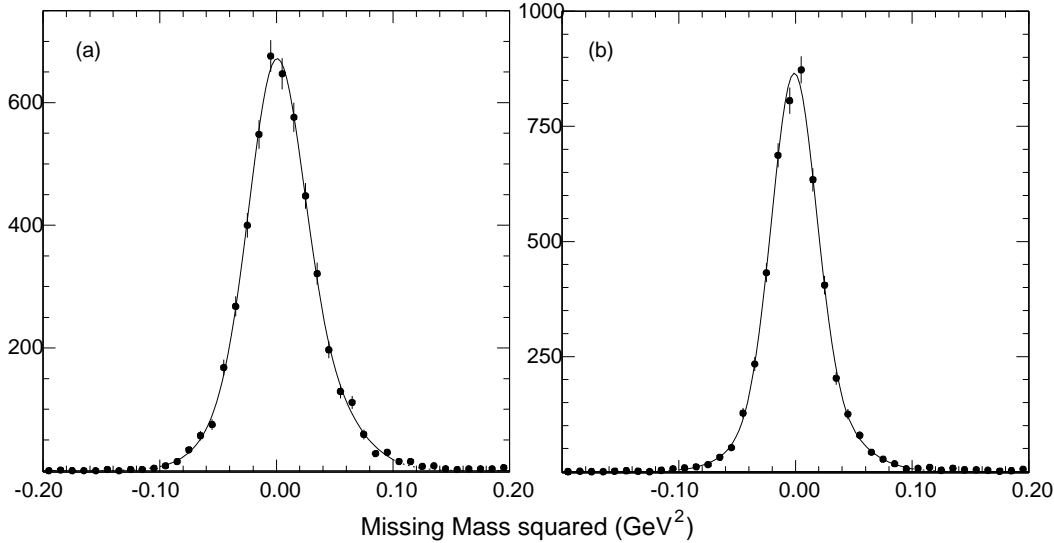}
\caption{The MM$^2$ resolution from Monte Carlo simulation for
$D_s^+\to\mu^+\nu$ utilizing a $\phi\pi^+$ tag and a $\gamma$ from
either $D_s^*$ decay, both before the kinematic fit (a) and after
(b).} \label{mc-munu-res}
\end{figure}

For the $\mu^+\nu$, final state the MM$^2$ distribution can be
modelled as the sum of two-Gaussians centered at zero (see
Eq.~\ref{eq:twoGauss}) A Monte Carlo simulation of the MM$^2$ for
the $\phi\pi^+$ subset of $K^+K^-\pi^+$ tags is shown in
Fig.~\ref{mc-munu-res} both before and after the fit. The fit
changes the resolution $\sigma$=0.032 GeV$^2$ to 0.025 GeV$^2$, a
22\% improvement, and there is no loss of events.

\begin{figure}[hbt]
\centering
\includegraphics[width=4in]{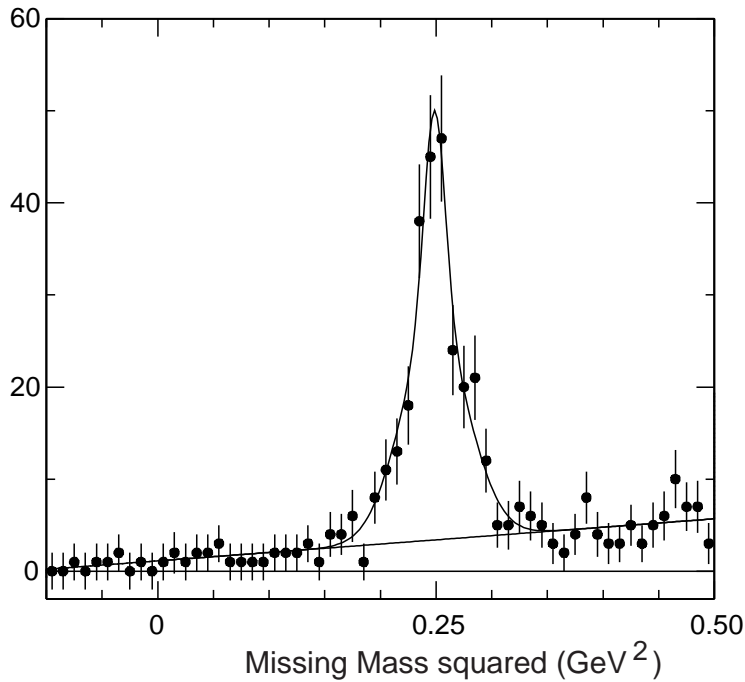}
\caption{The MM$^2$ distribution for events with a identified $K^+$
track. The kinematic fit has been applied. The curve is a fit to the
sum of two-Gaussians centered at the square of the $\overline{K}^0$
mass and a linear background.} \label{Kmm2-mc-data}
\end{figure}

We check the resolution using data. The mode $D_s^+\to K^0K^+$
provides an excellent testing ground. We search for events with at
least one additional track identified as kaon using the RICH
detector, in addition to a $D_s^-$ tag. The MM$^2$ distribution is
shown in Fig.~\ref{Kmm2-mc-data}. Fitting this distribution to a
two-Gaussian shape gives a MM$^2$ resolution of 0.025 GeV$^2$ in
agreement with Monte Carlo simulation.

\begin{figure}[hbt]
\centering
\includegraphics[width=3in]{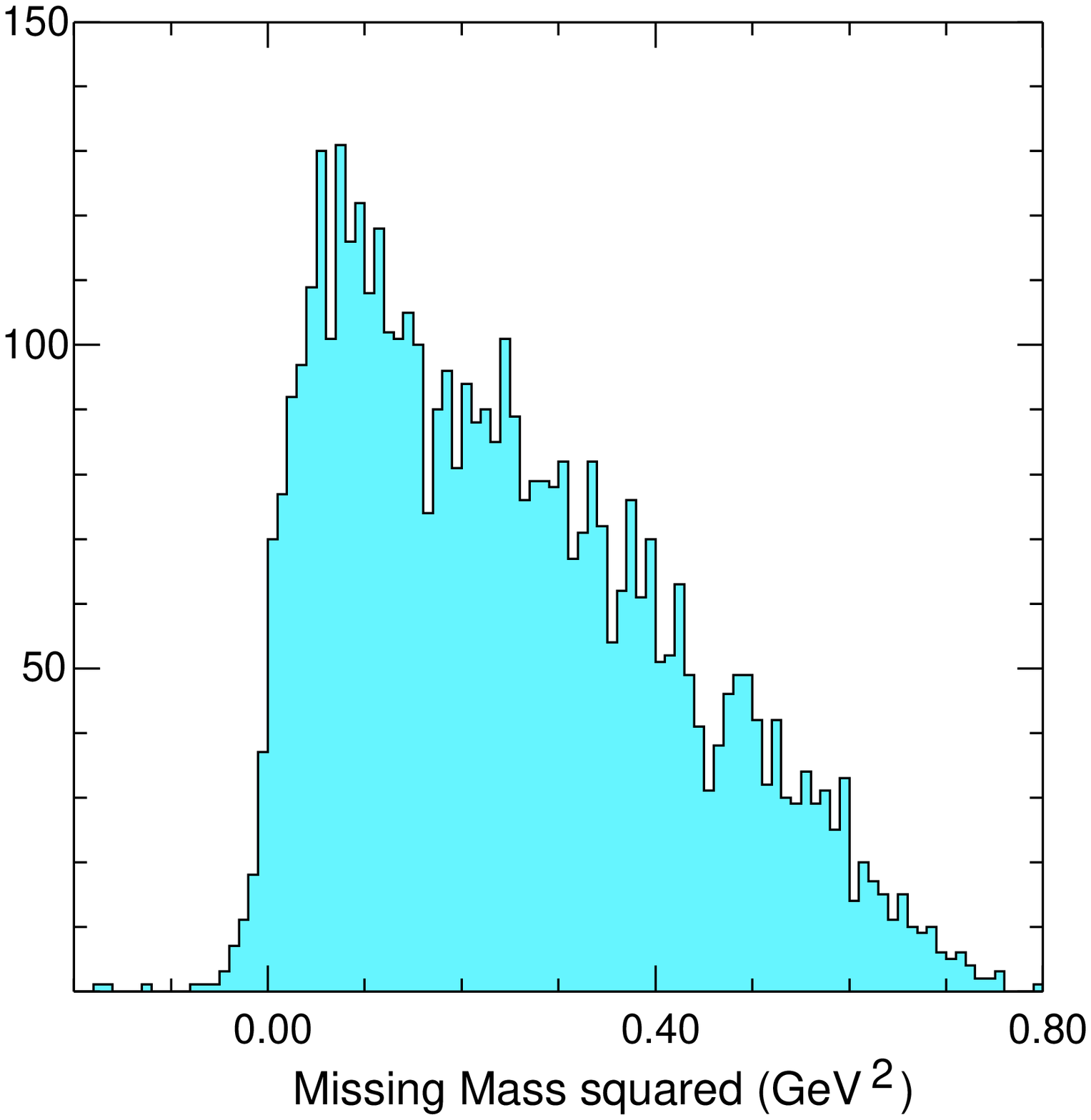}
\caption{The MM$^2$ distribution for $D_s^+\to\tau^+\nu$,
$\tau^+\to\pi^+\overline{\nu}$ at 4170 MeV.}
\label{mm2-taunu-pinu-mc}
\end{figure}

For the $\tau^+\nu$, $\tau^+\to\pi^+\nu$ final state a Monte Carlo
simulation of the MM$^2$ spectra is shown in
Fig.~\ref{mm2-taunu-pinu-mc}. The extra missing neutrino results in
a smeared distribution.

\subsection{MM$^2$ Spectra in Data}

\begin{figure}[htbp]
\centerline{ \epsfxsize=4.0in \epsffile{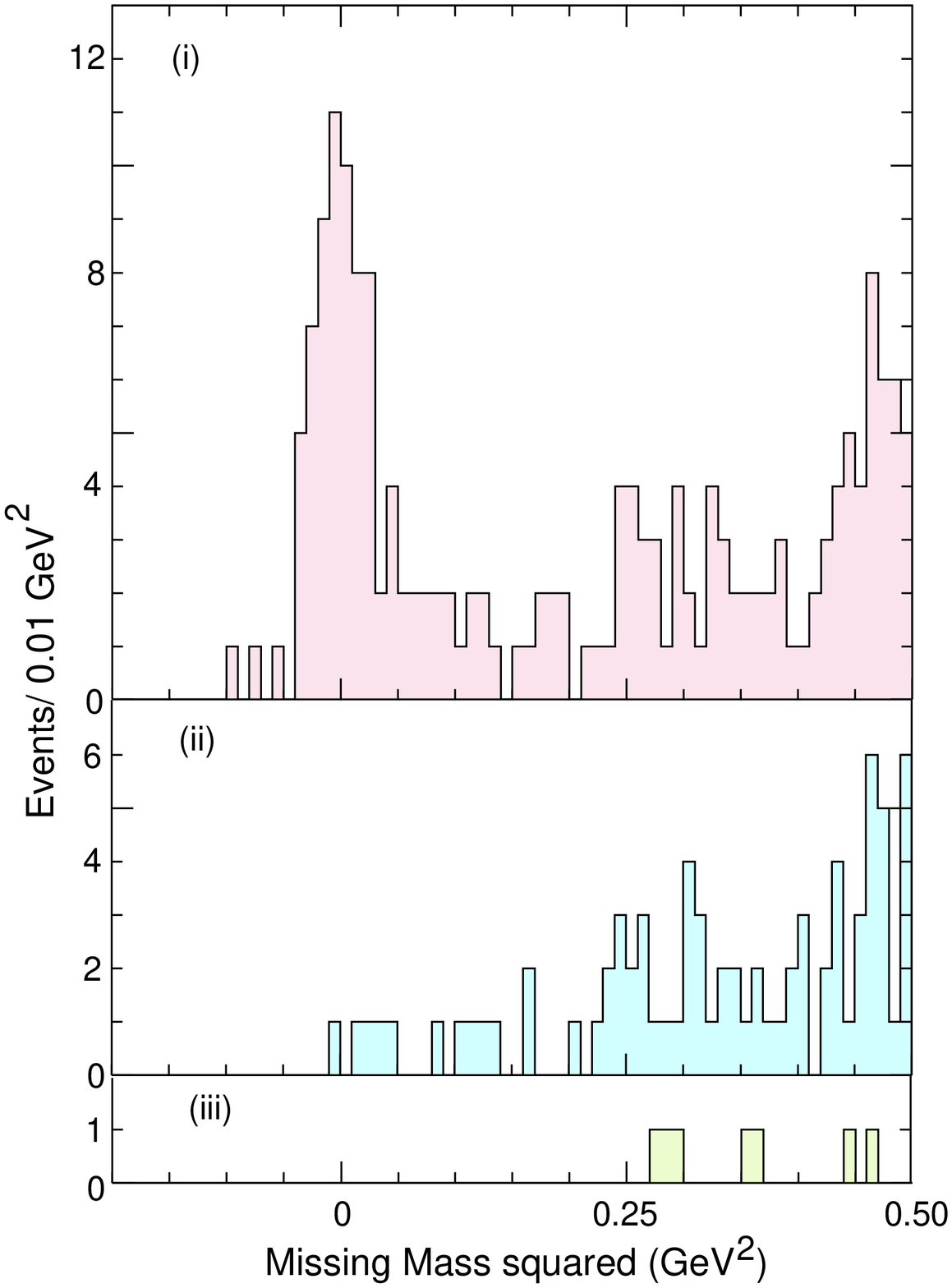}}
 \caption{The MM$^2$ distributions from data using $D_s^-$ tags and
 one additional opposite-sign
charged track and no extra energetic showers (see text). For the
case when the single track (a) deposits $<$~300 MeV of energy in the
calorimeter, case (i). The peak near zero is from $D_s^+\to\mu^+\nu$
events. (b) Track deposits $>$~300 MeV in crystal calorimeter but is
not consistent with being an electron, case (ii). (c) The track is
identified as an electron case (iii). } \label{mm2-data}
 \end{figure}
 The MM$^2$ distributions from data are shown in Fig.~\ref{mm2-data}.
Case (i) requires that the candidate muon track deposit $<$300 MeV
of energy in the calorimeter, consistent with a non-interacting
track. Case (ii) requires just the opposite, that the track deposit
$>$ 300 MeV and also not be consistent with being an electron. Case
(iii) requires that the track be consistent with being an electron.
The overall signal region we consider are below MM$^2$ of 0.20
GeV$^2$. Otherwise we admit background from $\eta\pi^+$ and
$K^0\pi^+$ final states. There is a clear peak in
Fig.~\ref{mm2-data}(a), due to $D_s^+\to\mu^+\nu$. Furthermore the
region between $\mu^+\nu$ peak and 0.20 GeV$^2$ has events that we
will show are dominantly due to the $\tau^+\nu$ decay.

The specific signal regions are defined as follows: for
$\mu^+\nu$, $0.05>$MM$^2>-0.05$ GeV$^2$, corresponding to $\pm
2\sigma$ or 95\% of the signal; for $\tau\nu$,
$\tau^+\to\pi^+\overline{\nu}$, in case (i) $0.20>$MM$^2>0.05$
GeV$^2$ and in case (ii) $0.20>$MM$^2>-0.05$ GeV$^2$. In these
regions we find 64, 24 and 12 events, respectively.

\subsection{Background Evaluations}

We consider the background arising from two sources: one from real
$D_s^+$ decays and the other from the background under the
single-tag signal-peaks. For the latter, we obtain the background
from data.
 We define side-bands of the invariant mass signals shown in
Fig.~\ref{mbc} in intervals approximately 4-5$\sigma$ on the low and
high sides of the invariant mass peaks for all modes. Thus the
amount of data corresponds to approximately twice the number of
background events under the signal peaks, except for the $\eta\pi^-$
and $\eta\rho^-$ modes, where the signal widths are so wide that we
chose narrower side-bands only equaling the data.  We analyze these
events exactly the same as those in the signal peak.

We list the backgrounds as the number from all modes but
$\eta\pi^-$ and $\eta\rho^-$ first, divided by 2 and the number
for these two modes. For case(i) we find 2/2+1 background in the
$\mu^+\nu$ signal region and 3/2+1 background in the $\tau^+\nu$
region. For case (ii) we find 2/2 events. Our total background
sample summing over all of these cases is 5.5$\pm$1.9.

 This entire procedure
was checked by doing the same study on a sample of Monte Carlo
generated at 4170 MeV that includes known charm and continuum
production cross-sections. The Monte Carlo sample is 7 times the
data. We find the number of background events to be predicted
directly by the simulation to be 28 and the sideband method yields
22. These numbers are slightly smaller that what is found in the
data but consistent within errors. We note that the Monte Carlo is
far from perfect in having many estimated branching fractions.

\begin{figure}[hbt]
\centering
\includegraphics[width=3in]{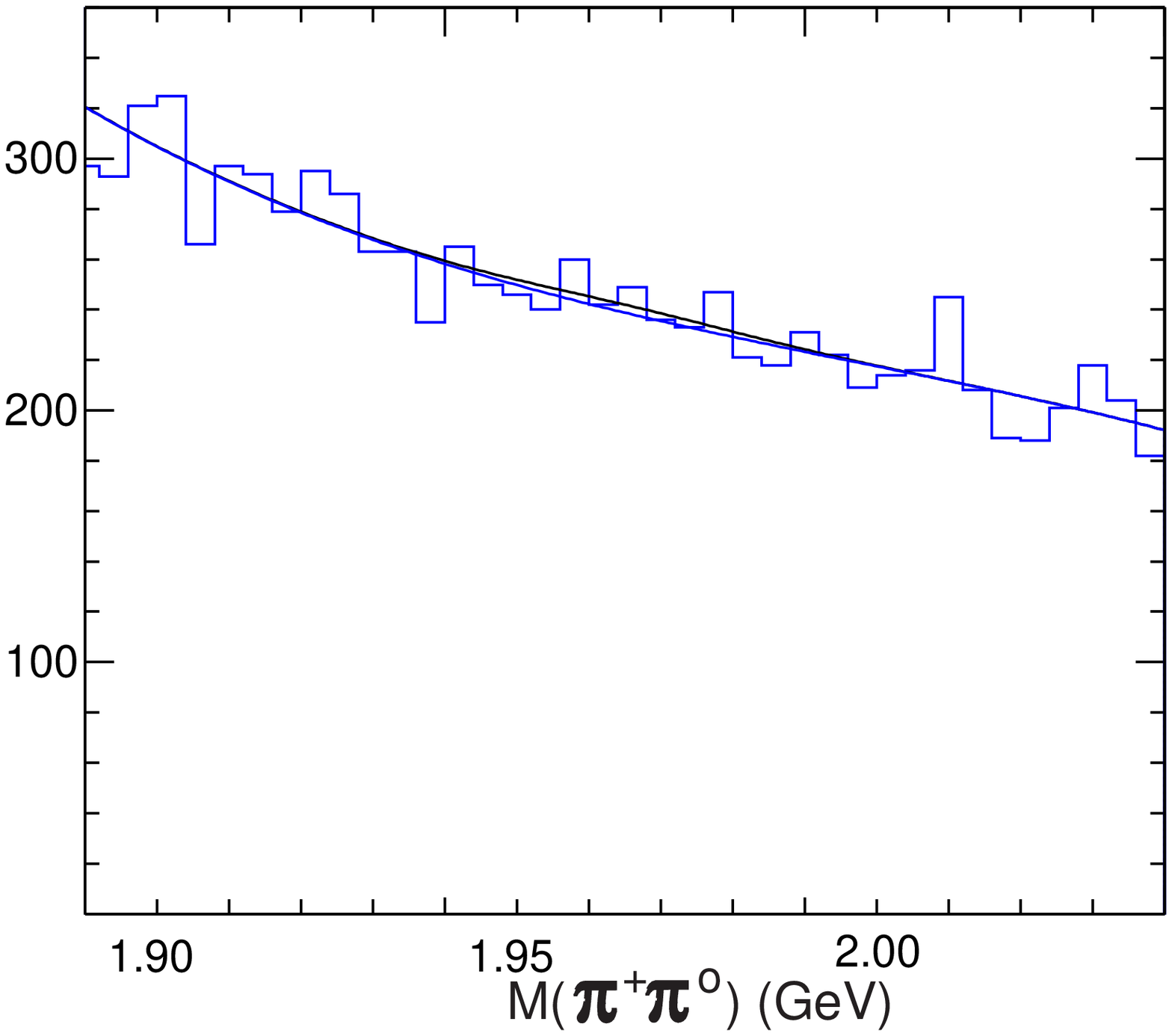}
\caption{The invariant $\pi^+\pi^0$ mass. The curves are fits to a
background polynomial and Gaussian signal with a width fixed by
Monte Carlo simulation. } \label{pipi0-single-tag}
\end{figure}

The background from real $D_s^+$ decays is studied by identifying
each possible source of background mode by mode. For the
$\mu^+\nu$ final state, the only possible background within the
signal region is $D_s^+\to\pi^+\pi^0$. This mode has not been
studied previously. We show in Fig.~\ref{pipi0-single-tag} the
$\pi^+\pi^0$ invariant mass spectrum. We don't see a signal and
set an upper limit $<1.1\times 10^{-3}$ at 90\% confidence level
(C. L.). Recall, that any such events are heavily suppressed by
the extra photon energy cut of 300 MeV. There are also some
$D_s^+\to\tau^+\nu$, $\tau^+\to\pi^+\overline{\nu}$ events that
occur in the signal region. We treat these as part of the signal
using the Standard Model expected ratio of decay rates from
Eq.~\ref{eq:tntomu} to calculate this contribution.

For the $\tau^+\nu$, $\tau^+\to\pi^+\overline{\nu}$ final state the
real $D_s^+$ backgrounds include, in addition to the $\pi^+\pi^0$
background discussed above, semileptonic decays, possible
$\pi^+\pi^0\pi^0$ decays and other $\tau^+$ decays. Semileptonic
decays involving muons are equal to those involving electrons shown
in Fig.~\ref{mm2-data}(c). Since no electron events appear the
signal region, the background from muons is also consistent with
zero. The $\pi^+\pi^0\pi^0$ background is estimated by considering
the $\pi^+\pi^+\pi^-$ final state whose measured branching ratio is
(1.02$\pm$0.12)\% \cite{stone-fpcp}. This mode has large
contributions from $f_0(980)\pi^+$ and other resonant structures in
$\pi^+\pi^-$ at higher mass \cite{focus-3pi}. The $\pi^+\pi^0\pi^0$
mode will also have these contributions, but the MM$^2$ opposite to
the $\pi^+$ will be at large mass. The only component that can
potentially cause background for us is the non-resonant part
measured by FOCUS as (17$\pm$4)\%. This background has been
evaluated by Monte Carlo simulation as well as others from other
$\tau^+$ decays, and listed in Table~\ref{tab:taunubkrd}.

\begin{table}[htb]
\begin{center}
\caption{Backgrounds in the $D_s^+\to\tau^+\nu$,
$\tau^+\to\pi^+\overline{\nu}$ sample, case (i) for
0.2$>$MM$^2>0.05$ GeV$^2$ and case (ii) for 0.2$>$MM$^2>-0.05$
GeV$^2$ \label{tab:taunubkrd}}
\begin{tabular}{lcccc}
 \hline\hline
    Mode  & ${\cal{B}}$(\%)            &  \# of events case(i) &  \# of events case(ii) & Sum\\ \hline
$\pi^+\pi^0\pi^0 $ & 1.0  & 0.1 & 0.3 & 0.4\\
$D_s^+\to\tau^+\nu$ & 6.2 & & &\\
~~~~$\tau^+\to \pi^+\pi^0\nu$ & 1.6 & 0.3 & 0.4 & 0.7\\
~~~~$\tau^+\to \mu^+\overline{\nu}\nu$ & 1.1 & 0 & 0 & 0\\
~~~~$\tau^+\to e^+\overline{\nu}\nu$ & 1.1& 0.2 & 0 & 0.2\\
\hline
Sum & & 0.6 & 0.7 & 1.3\\
\hline\hline
\end{tabular}
\end{center}
\end{table}

\section{Checks of the Method}

We perform an overall check of our procedures by measuring
${\cal{B}}(D_s^+\to \overline{K}^0K^+)$.  For this measurement we
compute the MM$^2$ (Eq.~\ref{eq:mm2}) using events with an
additional charged track but here identified as a kaon. These track
candidates have momenta of approximately 1 GeV/c; here our RICH
detector has a pion to kaon fake rate of 1.1\% with a kaon detection
efficiency of 88.5\% \cite{fakes}. For this study, we do not veto
events with extra charged tracks or photons.

The MM$^2$ distribution is shown in Fig.~\ref{Kmm2-mc-data}. The
peak near 0.25 GeV$^2$ is due to the decay mode of interest. We
fit this to a linear background from 0.02-0.50 GeV$^2$ plus a
two-Gaussian signal function. The fit yields 228$\pm$18$\pm$8
events. Since $\eta K^+$ could in principle contribute an
asymmetric background in this region, we searched for this final
state. Not finding any signal, we set an upper limit of 2.8$\times
10^{-3}$, approximately a factor of five below our measurement. In
order to compute the branching fraction we must include the
efficiency of detecting the kaon track 76.2\%, including radiation
\cite{gammamunu}, and take into account that it is easier to
detect the photon from the $D_s^*$ decay in $\overline{K}^0K^+$
events than in the average $D_sD_s^*$ event due to the track and
photon multiplicities, which gives a 4\% correction. These rates
are estimated by using Monte Carlo simulation. We determine
\begin{equation}
{\cal{B}}(D_s^+\to \overline{K}^0K^+)=(2.74\pm0.23\pm 0.13)\%,
\end{equation}
where the systematic errors are listed in Table~\ref{tab:sysKK}. The
largest components of the systematic errors arise from the number of
signal events (4.2\%) from the signal function width (3\%) and the
shape of the background function (3\%). This method is in good
agreement with the latest double tag result \cite{Peter}.

\begin{table}[htb]
\begin{center}

\caption{Systematic errors on determination of the $D_s^+\to
\overline{K}^0K^+$ branching ratio. \label{tab:sysKK}}
\begin{tabular}{lc} \hline\hline
   Error Source & Size (\%) \\ \hline
Signal shape & 1\\
Background shape & 1\\
Track finding &0.7 \\
PID cut &1.0 \\
Number of tags& 4.2\\
\hline
Total & 4.6 \\
 \hline\hline
\end{tabular}
\end{center}
\end{table}

\section{Leptonic Branching Fractions}

\begin{figure}[htbp]
 \vskip 0.00cm
\centerline{ \epsfxsize=4.0in \epsffile{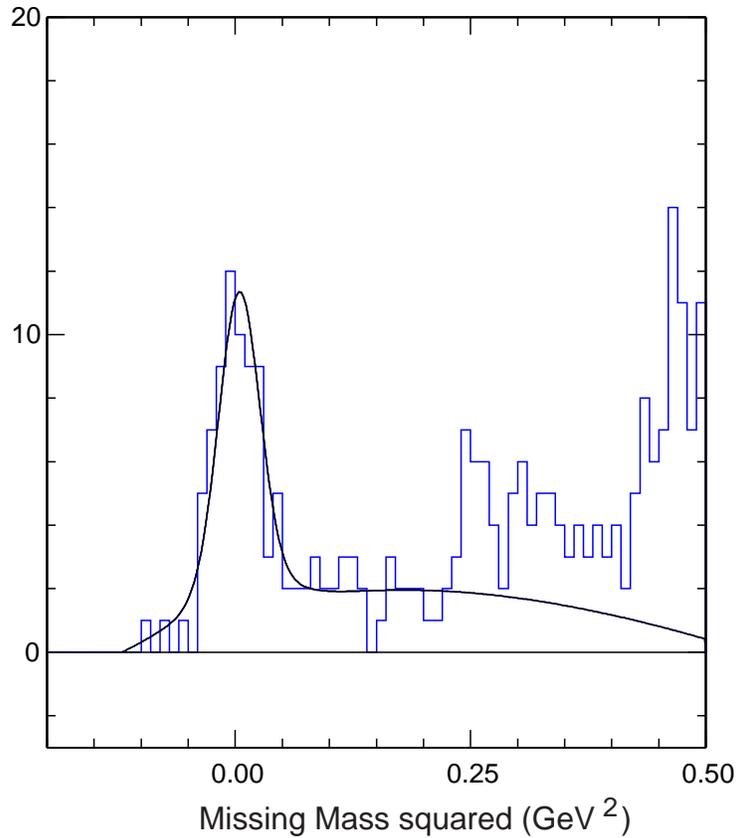} }
 \caption{The sum of case (i) and case (ii) MM$^2$ distributions compared to
 the predicted shapes for $D_s^+\to \mu^+\nu$ + $\tau^+\nu$, $\tau^+\to \pi^+\nu$.} \label{try1-total}
 \end{figure}

The sum of MM$^2$ distributions for case (i) and case (ii)
normalized to the expectation of the sum of $D_s^+\to \mu^+\nu$ and
$D_s^+\to \tau^+\nu$, $\tau^+\to \pi^+\nu$ is shown in
Fig.~\ref{try1-total}. The data are consistent with our expectation
containing mostly signal for MM$^2<$0.2 GeV$^2$. Recall there are
100 total events only 5.5 of which we estimate are background. Above
0.2 GeV$^2$ other, larger backgrounds enter.

The number of $\mu^+\nu$ events in the signal region,
$N_{\mu\nu}$, is related to the number of tags, $N_{\rm tag}$ the
branching ratios and the estimated backgrounds $N_{\rm bkgrd}$ as
\begin{equation}
 N_{\mu\nu}-N_{\rm bkgrd}=N_{\rm tag}\cdot
\epsilon\left[\epsilon'{\cal{B}}(D_s^+\to
\mu^+\nu)+\epsilon''{\cal{B}}(D_s^+\to
\tau^+\nu;~\tau^+\to\pi^+\nu)\right], \label{eq:munuB}
\end{equation}
where $\epsilon$ (79.5\%) includes the efficiency for
reconstructing the single charged track including final state
radiation (77.8\%), the (98.3$\pm$0.2)\% efficiency of not having
another unmatched cluster in the event with energy greater than
300 MeV, and for the fact that it is easier to find tags in
$\mu^+\nu$ events than in generic decays by 4\%, as determined by
Monte Carlo simulation. The efficiency labeled $\epsilon'$ is a
product of the 99\% muon efficiency for depositing less than 300
MeV in the calorimeter and 92.3\% acceptance of the MM$^2$ cut of
$|$MM$^2 < 0.05|$. The quantity $\epsilon''$ is the fraction of
$\tau^+\nu;~\tau^+\to\pi^+\nu$ events contained in the $\mu^+\nu$
signal window (13.2\%) times the 60\% acceptance for a pion to
deposit less than 300 MeV/c in the calorimeter (6.4\%).

The two $D_s^+$ branching ratios in Eq.~\ref{eq:munuB} are related
as
\begin{equation}
{\cal{B}}(D_s^+\to
\tau^+\nu;~\tau^+\to\pi^+\nu)=R\cdot{\cal{B}}(\tau^+\to\pi^+\nu){\cal{B}}(D_s^+\to
\mu^+\nu)=1.076\cdot{\cal{B}}(D_s^+\to \mu^+\nu)~,
\end{equation}
where we take the Standard Model ratio for $R$ as given in
Eq.~\ref{eq:tntomu} and ${\cal{B}}(\tau^+\to\pi^+\nu)$=11.06\%
\cite{PDG}. This allows us to solve Eq.~\ref{eq:munuB}. Since
$N_{\mu\nu}$= 64, $N_{\rm bkgrd}$=2, and
$N_{tag}=11880\pm399\pm499$, we find
\begin{equation}
{\cal{B}}(D_s^+\to \mu^+\nu)= (0.657\pm 0.090\pm0.028)\%.
\end{equation}

We can also sum the $\mu^+\nu$ and $\tau^+\nu$ contributions,
where we restrict ourselves to the MM$^2$ region below 0.20
GeV$^2$. Eq.~\ref{eq:munuB} still applies. The number of signal
and background events changes to 100 and 7.4, respectively. The
efficiency  $\epsilon'$ becomes unity, and $\epsilon''$ increases
to 45.2\%. Using this method, we find
\begin{equation}
{\cal{B}}(D_s^+\to \mu^+\nu)=(0.664\pm 0.076\pm0.028)\% .
\label{eq:finalBR}
\end{equation}

The systematic errors on these branching ratios are given in
Table~\ref{tab:munusys}.
\begin{table}[htb]
\begin{center}

\caption{Systematic errors on determination of the $D_s^+\to
\mu^+\nu$ branching ratio. \label{tab:munusys}}
\begin{tabular}{lc} \hline\hline
   Error Source & Size (\%) \\ \hline
Track finding &0.7 \\
Photon veto & 2 \\
Minimum ionization & 1\\
Number of tags& 4.2\\
\hline
Total & 5.2\\
 \hline\hline
\end{tabular}
\end{center}
\end{table}

We also can analyze the $\tau^+\nu$ final state independently. We
use different MM$^2$ regions for cases (i) and (ii) defined above.
For case (i) we define the signal region to be the interval
0.20$>$MM$^2 >$0.05
 GeV$^2$, while for case (ii) we define the signal region to be the interval
 0.20$>$MM$^2 >$-0.05 GeV$^2$. Case (i) includes 98\% of the $\mu^+\nu$ signal,
 so we must exclude the region close to zero MM$^2$, while for case (ii) we are specifically
 selecting pions so the signal region can be larger.
 The upper limit on MM$^2$ is
 chosen to avoid background from the tail of the $\overline{K}^0\pi^+$ peak.
 The fractions of the MM$^2$
 range accepted are 32\% and 45\% for case (i) and (ii), respectively.

 We find 24 signal and 3.1 background events for case (i) and 12 signal and 1.7 backgrounds
 for case (ii). The branching ratio, averaging the two cases is
\begin{equation}
{\cal{B}}(D_s^+\to \tau^+\nu)=(7.1\pm 1.4\pm0.3)\% .
\end{equation}
Lepton universality in the Standard Model requires that the ratio
$R$ from Eq.~\ref{eq:tntomu} be equal to a value of 9.72. We measure
\begin{equation}
R\equiv \frac{\Gamma(D_s^+\to \tau^+\nu)}{\Gamma(D_s^+\to
\mu^+\nu)}= 10.8\pm 2.6 \pm 0.2~. \label{eq:tntomu2}
\end{equation}
Thus we see no deviation from the predicted value. Current results
on $D^+$ leptonic decays also show no deviations
\cite{ourDptotaunu}. The absence of any detected electrons
opposite to our tags allows us to set an upper limit of
\begin{equation}
{\cal{B}}(D_s^+\to e^+\nu)= 3.1\times 10^{-4}
\end{equation}
at 90\% confidence level; this is also consistent with Standard
Model predictions and lepton universality.

 Using our most precise value for ${\cal{B}}(D_s^+\to \mu^+\nu)$
 from Eq.~\ref{eq:finalBR}, that is derived using both our $\mu^+\nu$ and $\tau^+\nu$
 samples,
 and Eq.~\ref{eq:equ_rate}, we extract
 \begin{equation}
 f_{D_s}=282\pm 16 \pm 7 {~\rm MeV}.
 \end{equation}

We combine with our previous result \cite{our-fDp}
\begin{equation}
f_{D}^+=222.6\pm 16.7^{+2.8}_{-3.4}{\rm ~MeV.}
\end{equation}
 and find a preliminary value for
\begin{equation}
\displaystyle{\frac{f_{D_s^+}}{f_{D^+}}=1.27\pm 0.12\pm 0.03},
\end{equation}
where a small part of the systematic cancels in our two
measurements.

\section{Conclusions}
Theoretical models that predict $f_{D_s^+}$ and the ratio
$\frac{f_{D_s^+}}{f_{D^+}}$ are listed in Table~\ref{tab:Models}.
Our result is higher than most theoretical expectations. We are
consistent with Lattice-Gauge theory, and most other models, for the
ratio of decay constants.

\begin{table}[htb]
\begin{center}

\caption{Theoretical predictions of $f_{D^+}$ and
$f_{D_S^+}/f_{D^+}$. QL indicates quenched lattice calculations.}
\label{tab:Models}
\begin{tabular}{lccl} \hline\hline
    Model &$f_{D_s^+}$ (MeV) &  $f_{D^+}$ (MeV)          &  ~~~~~$f_{D_s^+}/f_{D^+}$           \\\hline
Lattice ($n_f$=2+1)  \cite{Lat:Milc} &
 $249 \pm 3 \pm 16 $&$201\pm 3 \pm 17 $&$1.24\pm 0.01\pm 0.07$ \\
QL (Taiwan) \cite{Lat:Taiwan} &
$266\pm 10 \pm 18$ &$235 \pm 8\pm 14 $&$1.13\pm 0.03\pm 0.05$ \\
QL (UKQCD) \cite{Lat:UKQCD}&$236\pm 8^{+17}_{-14}$ & $210\pm 10^{+17}_{-16}$ & $1.13\pm 0.02^{+0.04}_{-0.02}$\\
QL \cite{Lat:Damir} & $231\pm 12^{+6}_{-1}$&$211\pm 14^{+2}_{-12}$ &
$1.10\pm 0.02$\\
QCD Sum Rules \cite{Bordes} & $205\pm 22$ & $177\pm 21$ & $1.16\pm
0.01\pm 0.03$\\
QCD Sum Rules \cite{Chiral} & $235\pm 24$&$203\pm 20$ & $1.15\pm 0.04$ \\
Quark Model \cite{Quarkmodel}&268 &$234$  & 1.15 \\
Quark Model \cite{QMII}&248$\pm$27 &$230\pm$25  & 1.08$\pm$0.01 \\
Potential Model \cite{Equations} & 241& 238  & 1.01 \\
Isospin Splittings \cite{Isospin} & & $262\pm 29$ & \\
\hline\hline
\end{tabular}
\end{center}
\end{table}

By using a theoretical prediction for $f_{D_s^+}/f_{D^+}$ we can
derive a value for the ratio of CKM elements $|V_{cd}/V_{cs}|$.
Taking the value from Ref.~\cite{Lat:Milc} of $1.24\pm 0.01\pm
0.07$, we find
\begin{equation}
|V_{cd}/V_{cs}|=0.22\pm 0.03~,
\end{equation}
where the theoretical and experimental errors have been added in
quadrature. This value is consistent with expectations.

 We now compare our preliminary results with previous measurements. The
branching fractions modes and derived values of $f_{D_s^+}$ are
listed in Table~\ref{tab:fDs}. Our values are shown first. We are
generally consistent with previous measurements that are also higher
than the theory.

\begin{table}[htb]
\begin{center}

\caption{These results compared with previous measurements. Results
have been updated for new values of the $D_s$ lifetime. ALEPH uses
both measurements to derive a value for the decay
constant.\label{tab:fDs}}
\begin{tabular}{llccc}\hline\hline
Exp. & Mode  & ${\cal{B}}$& ${\cal{B}}_{\phi\pi}$ (\%) & $f_{D_s^+}$ (MeV) \\
\hline CLEO-c & $\mu^+\nu$ & $(6.57\pm 0.90\pm 0.34)10^{-3}$ & & $281\pm 19\pm 7$\\
CLEO-c & $\tau^+\nu$ & $(7.1\pm 1.4\pm 0.3)10^{-2}$&& $296\pm 29 \pm 7 $ \\
CLEO-c & combined & -& & $282\pm 16 \pm 7$  \\
CLEO \cite{CLEO}& $\mu^+\nu$ &$(6.2\pm 0.8\pm 1.3 \pm
1.6)10^{-3}$&
3.6$\pm$0.9&$273\pm19\pm27\pm33$\\
BEATRICE \cite{BEAT} & $\mu^+\nu$ &$(8.3\pm 2.3\pm 0.6 \pm
2.1)10^{-3}$& 3.6$\pm$0.9&$315\pm43\pm12 \pm39$\\
ALEPH \cite{ALEPH}& $\mu^+\nu$ &$(6.8\pm 1.1\pm 1.8)10^{-3}$ & 3.6$\pm$0.9& $285\pm 19\pm 40$ \\
ALEPH \cite{ALEPH}& $\tau^+\nu$ &$(5.8\pm 0.8\pm 1.8)10^{-2}$ & &  \\
OPAL \cite{OPAL} & $\tau^+\nu$ & $(7.0\pm 2.1 \pm 2.0)10^{-3}$ & & $286\pm 44\pm 41$  \\
L3 \cite{L3} &$\tau^+\nu$ & $(7.4\pm 2.8 \pm 1.6\pm 1.8)10^{-3}$ & & $302\pm 57\pm 32 \pm 37$  \\
BaBar \cite{Babar-munu} & $\mu^+\nu$& $(6.5\pm 0.8\pm 0.3 \pm
0.9)10^{-3}$ & 4.8$\pm$0.5$\pm$0.4 &
 $279\pm 17 \pm 6 \pm 19$\\
\\\hline\hline
\end{tabular}
\end{center}

\end{table}

 Most measurements
of $D_s^+\to\ell^+\nu$ are normalized with respect to ${\cal{B}}
(D_s^+\to\phi\pi^+)$.
 One measurement that isn't is that of OPAL, which
normalizes to a $D_s$ fraction in $Z^0$ events that is derived
from an overall fit to heavy flavor data at LEP \cite{HFAG}. It
still, however, relies on absolute branching fractions that are
hidden by this procedure, and the estimated error on the
normalization is somewhat smaller than that indicated by the error
on ${\cal{B}}_{\phi\pi}$ available at the time of their
publication. The L3 measurement is normalized to using a
calculation that the fraction of $D_s$ mesons produced in $c$
quark fragmentation is 0.11$\pm$0.02 and that ratio of $D_s^*/D_s$
production is 0.65$\pm$0.10. The ALEPH results use
${\cal{B}}_{\phi\pi}$ for their $\mu^+\nu$ results and a similar
procedure as OPAL for their $\tau^+\nu$ results. We note that the
recent BaBar result uses a larger ${\cal{B}}_{\phi\pi}$ than the
other results.

The CLEO-c determination of $f_{D_s^+}$ is the most accurate to
date. It also does not rely on the independent determination of
any normalization mode.

\section{Acknowledgments}

We gratefully acknowledge the effort of the CESR staff in providing
us with excellent luminosity and running conditions. This work was
supported by the National Science Foundation, the U.S. Department of
Energy, and the Natural Sciences and Engineering Research Council of
Canada.


\begin{thebibliography}{99}
\bibitem{Peter}
The most recent value for ${\cal{B}}(D_s^+\to K_SK^+)$ is
(1.50$\pm$0.08)\%, which when doubled gives (3.00$\pm$0.16)\%.
Private communication from Peter Onyisi.
\end{thebibliography}
\end{document}